\documentclass[conference]{IEEEtran}
\IEEEoverridecommandlockouts
\usepackage{cite}
\usepackage{amsmath,amssymb,amsfonts}
\usepackage{algorithmic}
\usepackage{graphicx}
\usepackage{textcomp}
\usepackage{xcolor}
\def\BibTeX{{\rm B\kern-.05em{\sc i\kern-.025em b}\kern-.08em
    T\kern-.1667em\lower.7ex\hbox{E}\kern-.125emX}}
    
\usepackage{booktabs} 
\usepackage{multirow}
\usepackage{float}
\usepackage{subfig}
\usepackage{xspace}
\usepackage{balance}
\usepackage[multiple]{footmisc}
\usepackage{hyperref}
\usepackage{booktabs}
\usepackage{tabularx}
\usepackage{caption}
\usepackage[english]{babel}
\usepackage[autostyle]{csquotes}
\usepackage[newcommands]{ragged2e}
\usepackage{flushend}

\usepackage[english]{babel}
\usepackage{
  graphicx,                     
  xspace                        
}

\newcommand{\iot}{IoT\xspace}

\begin{document}

\title{SoK: Beyond IoT MUD Deployments - Challenges and Future Directions}
%
%
%
%
%
%

\author{
\IEEEauthorblockN{Angelo Feraudo\IEEEauthorrefmark{1}, Poonam Yadav\IEEEauthorrefmark{2}, Richard Mortier\IEEEauthorrefmark{3}, Paolo Bellavista\IEEEauthorrefmark{1} and Jon Crowcroft\IEEEauthorrefmark{3}}

\IEEEauthorblockA{\IEEEauthorrefmark{1}
Department of Computer Science and Engineering, 
University of Bologna, Italy}
\IEEEauthorblockA{\IEEEauthorrefmark{2}Department of Computer Science, 
University of York, UK, 
e-mail: poonam.yadav@york.ac.uk}
\IEEEauthorblockA{\IEEEauthorrefmark{3} Computer Lab,
University of Cambridge, UK}
}

\maketitle

\begin{abstract}
Due to the advancement of IoT devices in both domestic and industrial environments, the need to incorporate a mechanism to build accountability in the IoT ecosystem is paramount. In the last few years, various initiatives have been started in this direction addressing many socio-technical concerns and challenges to build an accountable system. The solution that has received a lot of attention in both industry and academia is the Manufacturer Usage Description (MUD) specification. It gives the possibility to the IoT device manufacturers to describe communications needed by each device to work properly. MUD implementation is challenging not only due to the diversity of IoT devices and manufacturer/operator/regulators but also due to the incremental integration of MUD-based flow control in the already existing Internet infrastructure. To provide a better understanding of these challenges, in this work, we explore and investigate the prototypes of three implementations proposed by different research teams and organisations, useful for the community to understand which are the various features implemented by the existing technologies. By considering that there exist some behaviours which can be only defined by local policy, we propose a MUD capable network integrating our User Policy Server (UPS). The UPS provides network administrators and end-users an opportunity to interact with MUD components through a user-friendly interface. Hence, we present a comprehensive survey of the challenges and future directions of MUD enabled environments.
\end{abstract}


\section{Introduction}
Due to the advancement and adoption of smart \iot systems in recent years, billions of sensors and actuator devices from thousands of vendors are connected to the Internet. The data generated from these devices generally have spatial and temporal dimensions, which have more contextual value to their local users than remote users.  However, in many use-cases, the collective data generated by distributed sensors/actuators bring value to their remote users. Furthermore,  the standard Internet protocols and Infrastructures are used to meet the remote access and connectivity requirements. The Internet connectivity not only speeds up the process of advancing new global and remote services, but also make these devices more vulnerable to new security, data, and privacy attacks similar to traditional internet-connected machines. Therefore, to minimise possible threats, \iot deployments need to address various security requirements: authentication, confidentiality, integrity, non-repudiation and availability.

Nevertheless, due to resource constraints and heterogeneity of current IoT devices, deployment of full-fledged security solutions is not possible.  Therefore, smart and distributed solutions are needed. One potential solution is to restrict the traffic communication pattern for these devices so that the attack surface is reduced. In order to simplify things and to bring progress in this direction, manufacturers of \iot devices should take responsibility and authority to provide legibility of the traffic generating from their IoT devices, and end-users, who are also network managers in domestic settings, should have user-friendly tools to analyse, detect and control over their IoT devices. 

Giving manufacturers the responsibility to provide a detailed description of the behaviours of the device, has been investigated~\cite{Cisco2019} and an initiative to formalise specifications by IETF (RFC 8520)~\cite{rfc8520}.   In our work, we present a summary of the current work in this direction and identify many research questions and challenges that need future investigation. Furthermore, we present our solutions to address some of these challenges by focussing on the deployment of manufacturer-provided MUD along with user-defined policies for COTS devices.

\textbf{Contributions:} In this paper, we analyse state-of-the-art MUD deployment scenarios and discuss challenges and shortcoming with each approach. We present the User Policy Server (UPS),  which provides to network administrators and end-users the opportunity to interact with MUD components through a user-friendly interface. However, the proof-of-concept provided presents new challenges that are analysed throughout this paper. Hence, we present a list of future work which we think is a contribution to the research community who are working in this direction.

The rest of the paper is organised as follows. In Section~\ref{section:overview}, we describe an overview of the MUD workflow in a home environment and explain how MUD can isolate the IoT devices from potential external and internal security attacks and provide short descriptions of three MUD deployments. In Section~\ref{findings},  we present current open challenges and their security impacts from our knowledge gained by reviewing the three MUD deployments.  In Section, \ref{challenges}, \ref{oursolutions} and \ref {evaluations}, we list some of the security challenges which could be solved by our solutions and then present our solutions and evaluations in the respective Sections. The related work in this direction so far and future work we discuss in Section~\ref{related} and  \ref{conclusion} respectively.
\section{MUD overview}\label{section:overview}
\begin{figure}[H]
    \centering
    \includegraphics[width=0.3\textwidth]{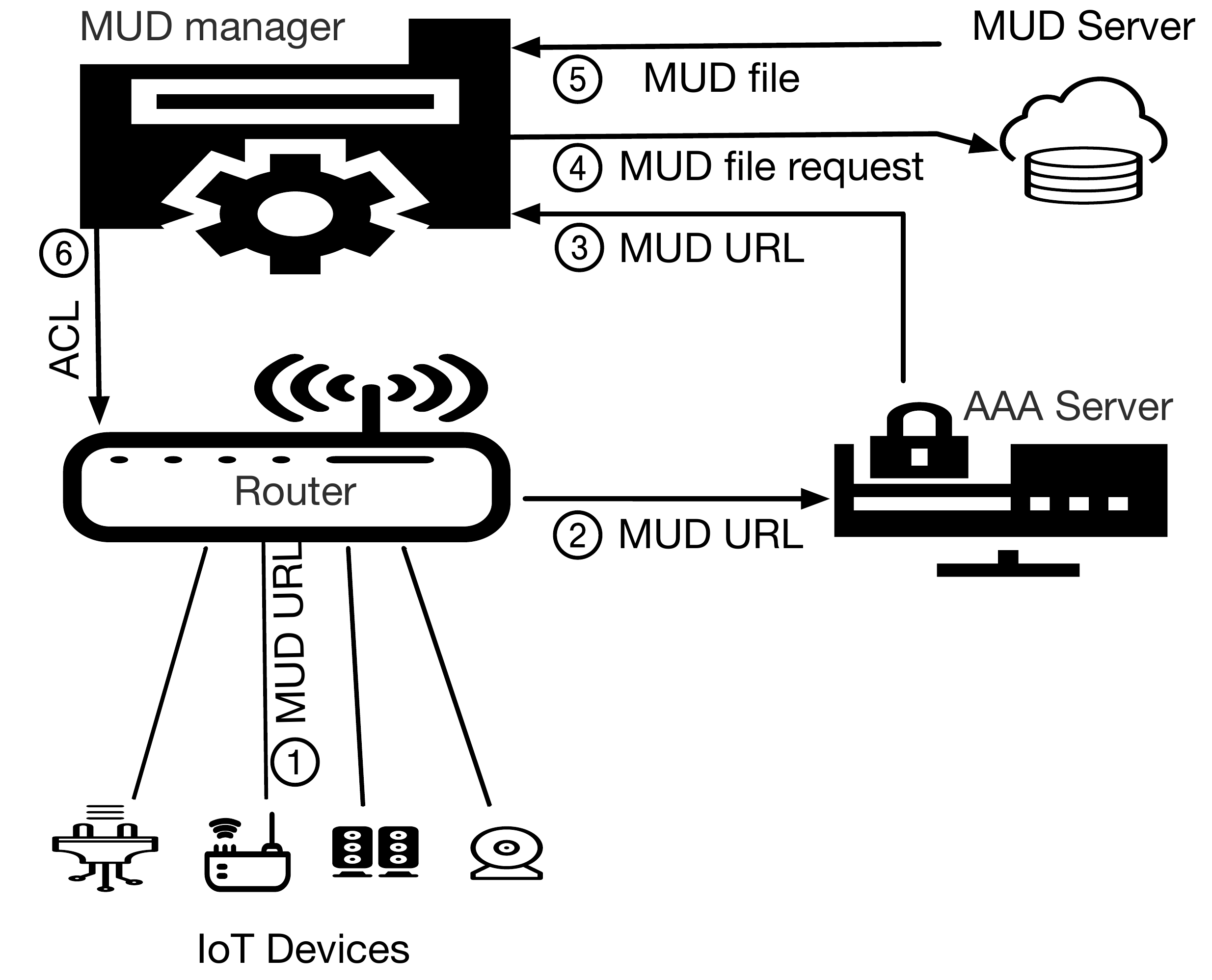}
    \caption{MUD Workflow.}
    \label{fig:Mudflow}
\end{figure}

According to IETF (RFC 8520)~\cite{rfc8520}, a MUD deployment should consist of three architectural building blocks: (1) The Manufacturer User Description (MUD) file, that is created by the device manufacturer to describe the device and its expected network behaviour. (2) A uniform resource locator (URL), that is used to retrieve the MUD file from a manufacturer's server when a device is added to a home or a small business network. (3)  A mechanism for local network management systems to retrieve the description. The workflow between these components is shown in Fig~\ref{fig:Mudflow}.   In \textcircled{1},  when a device added in the network, it sends MUD-URL with a X.509 certificate or DHCP or LLDP (depending on the implementation).  In \textcircled{2}, the local router  (As Network Access Device (NAD)) sends the MUD-URL to AAA server for authentication of the request. In \textcircled{3}, AAA server sends the authenticated MUD-URL to MUD manager.  The MUD Manager, in case of X.509 authentication, validate the signature and request  (\textcircled{4}) and retrieve MUD file from the manufacturer's file server (\textcircled{5}). The MUD file is a YANG-based JSON file (IETF RFC 7951)~\cite{rfc7951}  signed with a public key signature from the device manufacturer. The MUD manager validates and processes the MUD file and store file locally for until its certificate is valid and generate Access Control Lists (ACL). In \textcircled{6}, the MUD manager sends ACL to NAD. The NAD enforces ACL rules it received from the  MUD manager to all the traffic passing through it. 

From the workflow and the MUD specification, we can conclude that MUD specification has provided enough flexibility and scope to choose different implementations to incorporate requirements of various deployment scenarios. In our work,  we focus on investigating three different implementations of MUD deployments: Cisco proof-the-concept~(PoC) implementation, National Institute of Standards and Technology (NIST)~\cite{nist19, ranganathan2019soft} implementation based on Software Defined Network and  OpenSource MUD~\cite{osMud} developed by a consortium of companies (Cable Labs, Cisco, CTIA, Digicert, ForeScout, Global Cyber Alliance, Patton, and Symantec) in device manufacturing and network security, coordinated by a cybersecurity firm, MasterPeace Solutions. We present a comparison of these implementations in the Table~\ref{mudcompare} and provide brief descriptions in this section. 

\subsection{Cisco MUD PoC}
\begin{figure}[H]
    \centering
    \includegraphics[width=0.3\textwidth]{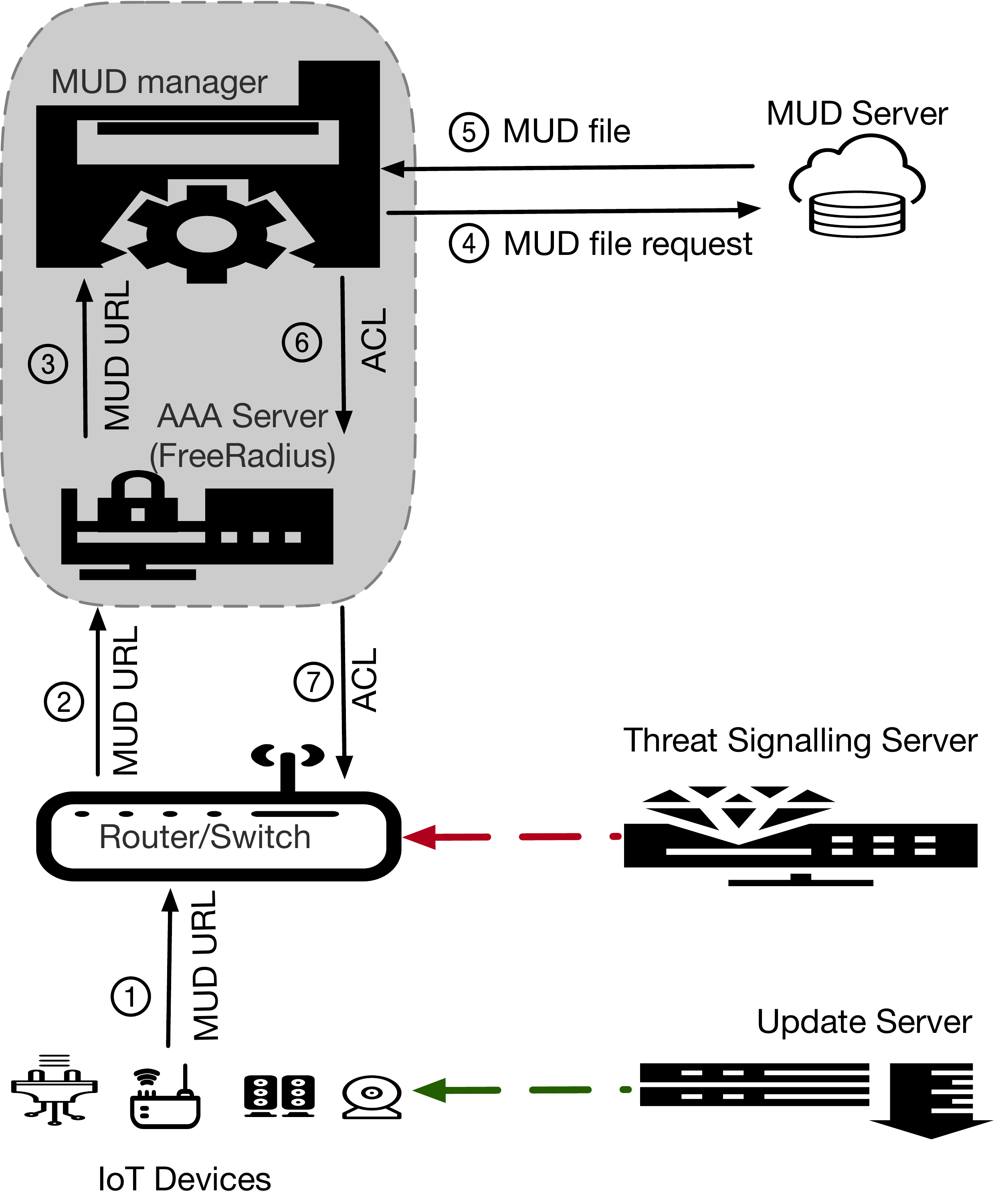}
    \caption{Logical Architecture of Cisco MUD implementation. }
    \label{fig:Cisco_MUD_workflow}
\end{figure}

Cisco proposed (shown in Figure~\ref{fig:Cisco_MUD_workflow})  an open-source proof-of-concept~(PoC) intended to help engineers and users to familiarise with the MUD standard. Therefore, the implementation does not include all the mechanisms typical of MUD environments. Some MUD manager automation is not implemented, and some protocol characteristics are not supported ~\cite{dodson2019securing}. 

As illustrated in Figure~\ref{fig:Cisco_MUD_workflow}, they provide a single device serving as MUD manager and a \textit{FreeRADIUS server} (as a AAA Server) that interfaces with the Catalyst 3850-S switch over TCP/IP.  The DHCP server hosted by the Catalyst 3850-S switch is configured to be MUD enabled, i.e. extract MUD URLs from IPv4 DHCP packets. The deployment provided leverages  DHCP and LLDP as MUD URL emission methods. The Cisco MUD Manager interacts with a FreeRADIUS server to authenticate the MUD URL received from the routers, and uses MongoDB to store policy information. Furthermore, its configuration shows the aim of this PoC, i.e. introducing MUD standards to manufacturer and end-users.
{}
Along with the standard MUD implementation, Cisco initial architecture also proposed two additional components, Threat Signalling Server and Update Server.  The requirements for DDoS Open Threat Signaling (DOTS) is described in IETF RFC 8612 ~\cite{rfc8612}.

\subsection{NIST MUD}\label{nistov}
The NIST implementation~\cite{nist19} relies on the Software Defined Networking concept, i.e. data plane and control plane disjunction, and OpenFlow SDN switches. In this deployment, the rules are organised in more than one flow tables by the OpenFlow switch. The SDN controllers manage the rules injection and update in two ways relying on particular times: dynamically, when a packet arrives at the controller, or proactively when the switch connects to the controller.  Ranganathan et al.~\cite{ranganathan2019soft} presented an implementation focusing on achieving rules scalability at SDN switch, by using SDN flow rules in three flow tables. The first two flow tables classify source and destination MAC addresses.
In comparison, the third table implements the MUD policy by introducing rules defined in terms of packet classification metadata assigned in the first two tables. After generating the pipeline described, the packet is finally sent to a table that implements L2Switch flow rules which is provided by another application. The latter forces the implementation to not use general Network ACEs.

\begin{figure}[H]
    \centering
    \includegraphics[width=0.4\textwidth]{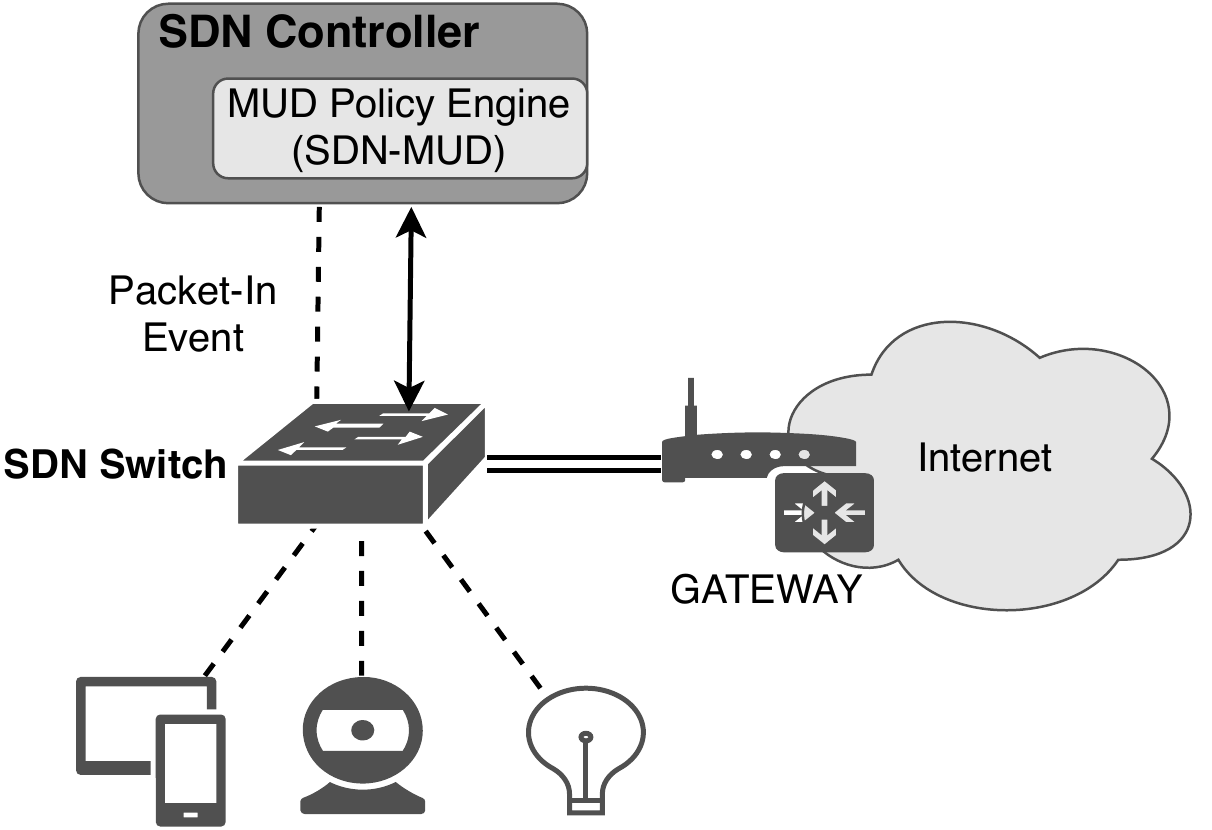}
    \caption{NIST-MUD Architecture.}
    \label{fig:nist-mudArchitecture}
\end{figure}

\subsection{Open Source MUD}
\begin{figure}[H]
    \centering
    \includegraphics[width=0.4\textwidth]{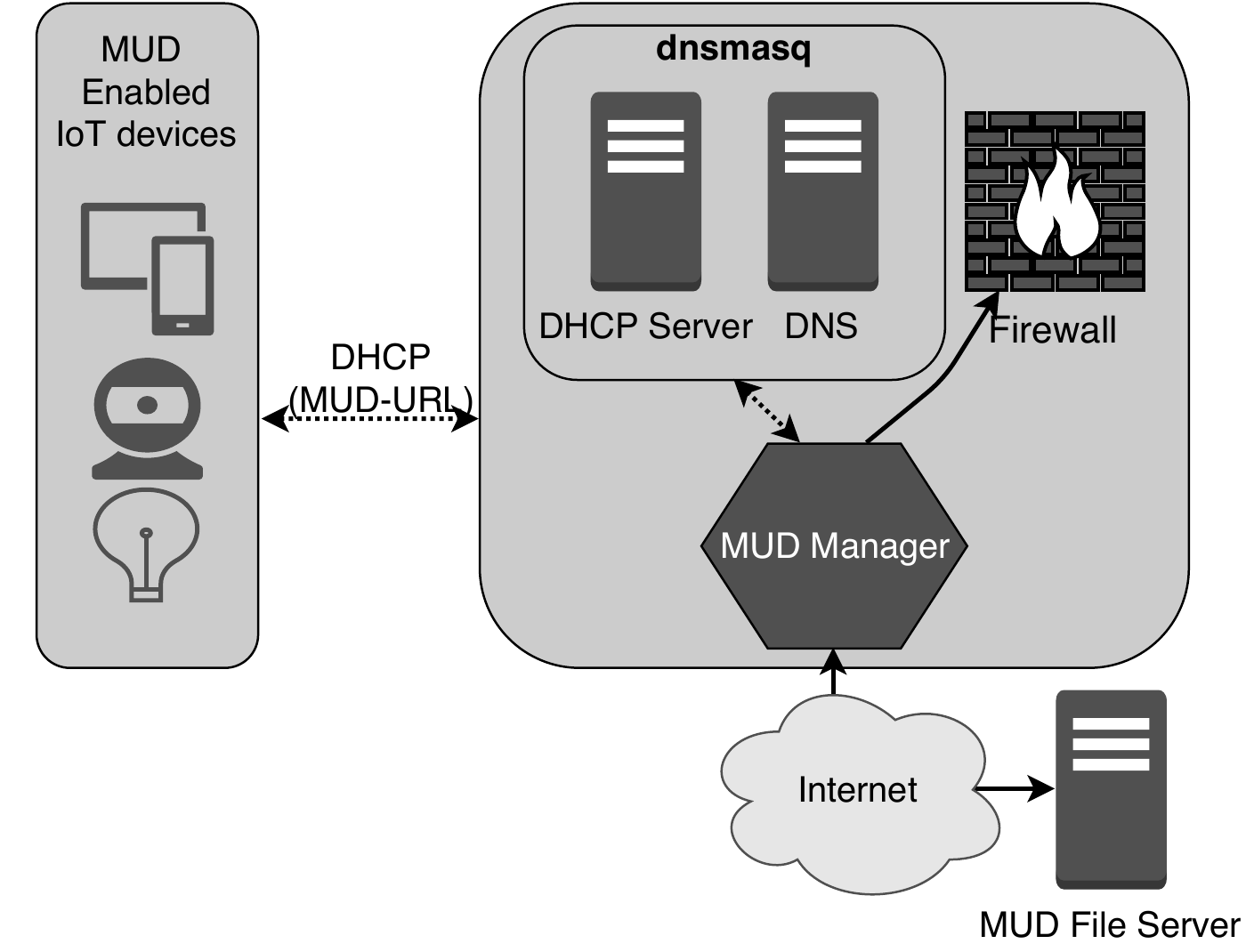}
    \caption{osMUD Architecture.}
    \label{fig:osMUDArchitecture}
\end{figure}

Open Source MUD (osMUD)~\cite{osMud} is an open-source implementation \cite{osMUDImpl} developed by a consortium of device manufacturing and network security companies. By comparing the architecture shown in Fig. \ref{fig:osMUDArchitecture} with those analysed in previous paragraphs, it is noticeable that the MUD manager directly runs on routers. Thus, only routers that have some characteristics are suitable for this deployment. OsMUD is designed easily build, deploy, and run on Open Wireless Router (OpenWRT) platform, which limits the router choice to only those compliant with this platform. Furthermore, the implementation integrates dnsmasq that extracts from the DHCP packets the MUD-URL, provides network infrastructure services, and is designed to be suitable for resource-constrained routers and firewalls.
       
\begin{table*}[t]
 \centering
 \captionof{table}{Comparison of MUD implementations\label{mudcompare}}
\resizebox{\textwidth}{!}{%
\begin{tabular}{cccccccc}
\toprule
\textbf{Implementation} & \textbf{Support} & \textbf{URL via DHCP} & \textbf{URL via LLDP} & \textbf{URL via x509} &  \textbf{URL via  another ways} & \textbf{Component of an AAA system}  & \textbf{Sig Verification}\\ \midrule

Cisco-MUD~& PoC - Cisco Catalyst 3850 & Yes (no DHCPv6) & Yes & No & No & Yes & OpenSSL\\ \\
NIST-MUD~& OpenVSwitch & Yes & No & No & with MAC + MUD & No& No\\ \\
osMUD~& Run on openwrt&dnsmasq& No&No&No&No&OpenSSL\\ 
\midrule
\end{tabular}
}
\end{table*}




\section{Findings: Limitations and Security issues}\label{findings}
In this section, we discuss the limitations and security issues of the three prototypes described in the previous section.  We investigated these implementations by setting a test environment in our lab. We understand these implementations are proofs-of-concept and are not fully implemented versions. Still, we thought this is a good starting point to understand the pros and cons of each implementation, which motivates further work in this context. Furthermore, this is the first existing analysis in this direction, which explores the challenges that must be considered during the building of a MUD capable network. 

\subsection{Cisco MUD }
As a proof-of-concept, the main limitations of this deployment regard the MUD Manager configuration. In fact, before the beginning of the MUD Manager, the configuration of DNS resolution service is manually performed by a human operator, which implies to restart the MUD manager process if new settings have been made. The static configuration introduces scalability problems, due to the manual configuration of DNS resolution, and availability problems, because the MUD manager needs to be restarted to pick the new settings made. Thus, Cisco deployment is impractical in real environments.

Another static limit, which regards the implementation itself, involves ACLs management. In particular, all rules, including ingress traffic, i.e. traffic received from external sources to the network and directed to local IoT devices, are static. Consequently, even MUD compliant IoT devices are still vulnerable to attacks originating from external domains, also though the device's MUD file makes it clear that the device is not authorised to receive traffic from that domain. Nevertheless, dynamic rules, including egress traffic, i.e. traffic sent from IoT devices to an external domain, are supported~\cite{dodson2019securing}. Thus, an attacker is not able to establish a TCP connection with the device from an outside domain.

\subsection{NIST MUD}
The SDN based PoC provided by Ranganathan et al. \cite{ranganathan2019soft} presents some issues on packet processing. By following the tables pipeline (section \ref{nistov}), the first rule in MAC address (first tables) classifications stage, which is installed when the switch connects, could allow the packet sending to the controller but not to the next table. Thus, a packet may not proceed in the tables pipeline before it can be classified, which implies performance consequences, i.e. switch failure because no packets from a newly connected device can go through the first table until the rule being installed. The behaviour described may be necessary if strict ACLs are required. However, Ranganathan et al. \cite{ranganathan2019soft}  proposed a  \enquote{relaxed} mode to manage these situations by loosening the ACE definition. In this mode packets can proceed in the pipeline while classification flow rules are being installed, which can result in the violation of the MUD ACEs with the condition that the system will become eventually compliant to the MUD ACEs. These packets could get through before the classification rule being installed at the switch.

\subsection{Open Source MUD}
The limits emerging from this implementation regards the deployment requirements. To host the MUD manager, only OpenWRT compliant routers using a particular version of dnsmasq can be employed in this deployment. A valuable solution provided by osMUD developers to run the MUD Manager outside OpenWRT is to compile it in C environments. However, it induces new requirements: it requires to use a compatible firewall and a DHCP server that can extract the MUD-URL from the DHCP header packet for MUD Enabled Devices. Thus, OpenWRT remains the most user-friendly alternative, as a result of availability for the majority of general-purpose routers \cite{openwrtSupported} and making the MUD manager deployment easier. 

Lastly, the current implementation does not have MUD file rules for lateral movement; thus, attackers can progressively move through a network, searching for targeted key data and assets.


\section{Security Challenges} \label{challenges}
In this section, we summarise the potential security implications of the three MUD implementations.  None of the implementations uses  X.509 option to emit the MUD URL and provides support through DHCP or LLDP.  Therefore, if a device is compromised, the MUD URL issued through it can potentially be spoofed.  Conversely, it is secure when the device uses the X.509 extension since the MUD URL is added to the certificate by the manufacturer. This means that the MUD URL emitted by an X.509 device can not be spoofed without detection, even if the device is exploited.
Further security issues are related to the MUD Specification itself. It does not provide any inherent security protections to IoT devices themselves. If a device's MUD file permits an IoT device to receive communications from a malicious domain, traffic from that domain can be used to attack the IoT device. Similarly, if the MUD file permits an IoT device to send communications to other domains, and if the IoT device is compromised, it can be used to attack those other domains. Some security considerations to keep in mind by users implementing MUD are provided by the specification \cite{rfc8520} in the \textit{Security Consideration} section. However, we list some additional security challenges which need further addressing:
(1) How to prevent a \iot device from a compromised \iot device's local attack in the local environment.
(2) When MUD signing certificate gets expires while MUD file is still deployed and in function locally; MUD Manager should fetch a newly signed file with signature and updates rules without interrupting the IoT device availability and functionality.
(3) MUD enforced rules does not provide sufficient fine-grained control. Individual users may have their deployment setup with may require additional rules; thus a mechanism to incorporate the user policy to filter traffic to provide more fine-grained control is needed.
(4) How to  incorporate updates from Threat Signalling Server. The further question and investigation in this direction is -  should they go through MUD file or through a generalised common server updates (and filters) for all IoT devices.
 
 \begin{figure}
    \centering
    \includegraphics[width=0.4\textwidth]{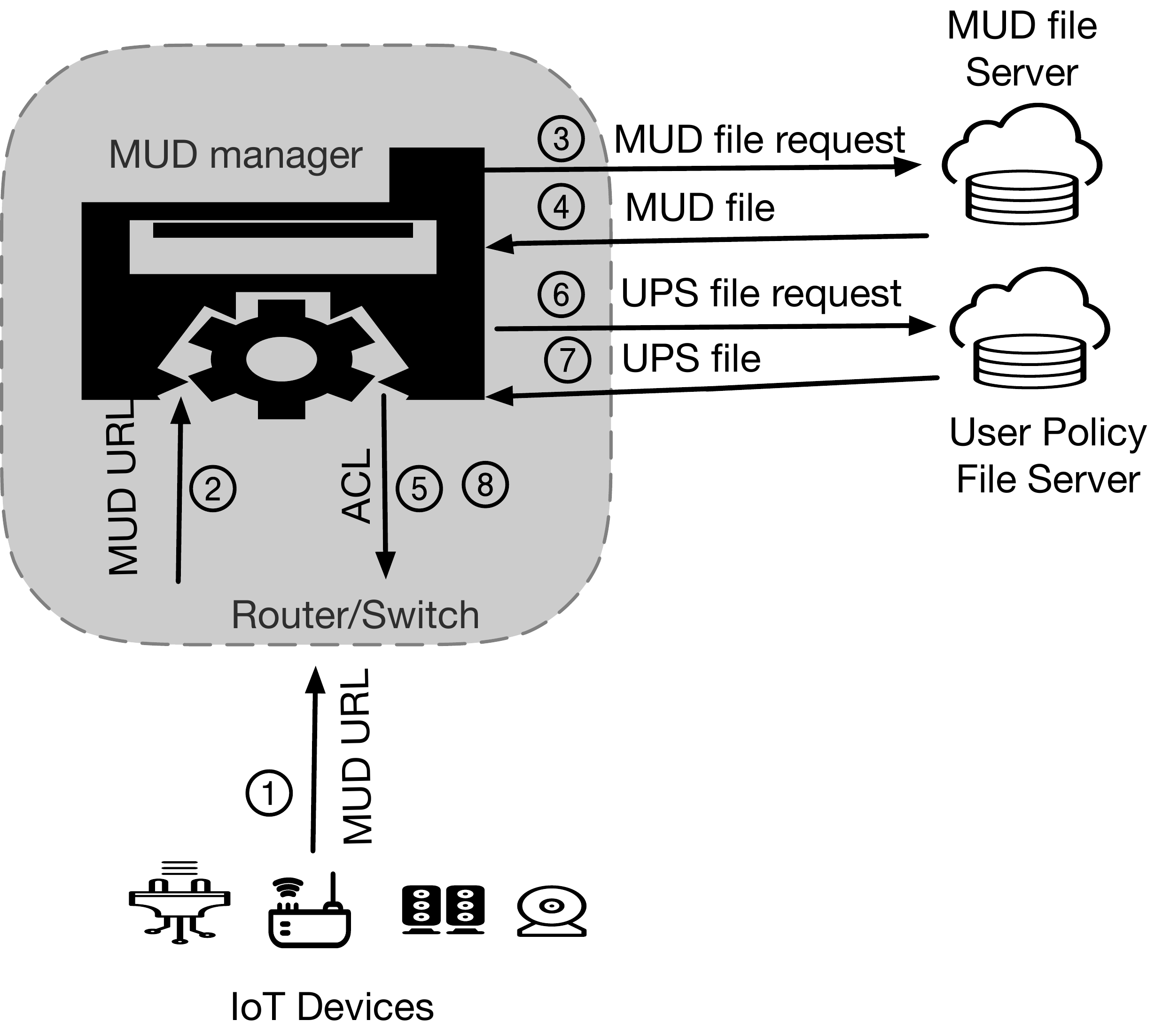}
    \caption{Our setup: osMUD with remote MUD file server and local User Policy Server.}
    \label{fig:osMUDlocal}
\end{figure}

\section{User Policy Server in a MUD enabled network}\label{oursolutions}
In this Section, we present and analyze our solution of problem three listed in the previous Section, which we consider of relative importance. The deployment showed in Figure~\ref{fig:osMUDlocal}, which was used in our testbed, has these main components: (1) OpenWRT router running osMUD Manager (2) Manufacturer MUD file server running remotely and (3) our User Policy Server (UPS) running locally on a local server.
The User Policy Server is a web server based on express framework whose architecture is similar to the Manufacturer MUD File Server (MFS). Deploying this entity gives to an administrator/end-user of an internal network the opportunity to store new MUD Files (hereafter called UPS Files) containing rules more suitable for the network in which MUD is deployed. Thus, a network administrator is now able to interact with all the MUD components using a user-friendly interface. 

It should be noted that by using the same YANG compliant \text{JSON} files and MFS interaction, the UPS does not imply relevant changes on the overall osMUD architecture.  In particular, the osMUD implementation is extended by us in order to incorporate the following workflow (Figure~\ref{fig:osMUDlocal}): (1) MUD Manager retrieves the MUD file from the manufacturer MUD file server; (2) MUD Manager  verify and parse files and generate ACL and insert ACL on Netgear router; (3) MUD Manager asks to the UPS if there is another UPS file for the device; (4) if yes, MUD Manager retrieves the file, verify and parse files and generate ACL and insert ACL on Netgear router. 

The previous steps imply that router hosting osMUD has UPS's certificate and that osMUD implementation can request a further UPS file. We assume that the MUD Manager knows a priori the UPS location. To locate UPS files and make the communication trusted, the device's mac-address is used as UPS file name(e.g. \textit{$<$device-mac-address$>$.json}) and each UPS file is signed by the UPS (as MFS case). After obtaining the UPS file, the MUD Manager can now process it and produce the rules to enforce in the router firewall. At the end of this procedure, the union of manufacturer and administrator rules is performed. This behaviour is appropriate to define either internal communication pattern or more relaxed rules. For example, in voice-based assistant systems (e.g., GoogleHome) case, it is hard for the manufacturer to define a communication pattern that allows all communications that it needs. Thus, the administrator can quickly build new communication patterns for the necessary services.

Nevertheless, the rules union could bring out new challenges: redundancy (limited storage) and conflicting rules (packet filtering performance \cite{firewall}). The former can be easily managed by using a matching technique, e.g. rule to insert with IP tables rules. The latter induces a vital problem that needs to be solved. For example, suppose that the manufacturer's rules (MUD file) allow the communication with two different external hosts, say, ${A}$ and ${B}$, but administrator rules (UPS file) allow only ${A}$ and ${C}$. In our current system, the device for which these rules are defined is allowed to communicate with all three hosts (${A}$, ${B}$ and ${C}$). If a priority system is used and administrator rules have the highest priority, the previous device is able to communicate with only ${A}$ and ${C}$. Both problems are beyond the aim of this work, but they will be considered for future UPS versions. Hence, considering the ability to remove/reversing MUD rules, using UPS needs thoughtful consideration. 

The UPS introduction in a MUD enabled the network to pave the way to new exciting scenarios:

\begin{itemize}
    \item \textbf{Automatic UPS File generation}: Use a tool that automatically generates UPS files (YANG compliant \text{JSON} files) by monitoring the internal traffic ~\cite{MUDGEE}.
    \item \textbf{New MUD File structure for UPS Files}: Extension of current file structure to obtain more restrictive rules such as packet rate, maximum packets, time restrictions, maximum ingress and egress points. To implement this, we require extending osMUD Manager, parsing capabilities, and rule enforcement logic.
    \item \textbf{Publish/subscribe architecture}: it is possible to extend the UPS infrastructure with a publish/subscribe model~\cite{Feraudo2020}, where server becomes the publisher and the router becomes subscriber, in order real time rules upgrade. The server can publish, by understanding the traffic behaviour, new rules that must be inserted in the IP tables of the router. The described architecture requires signature based authentication between the router and server, otherwise an attacker could impersonate the server and insert less restrictive rules.
\end{itemize}

\subsection{Evaluations}\label{evaluations}
The system evaluations aim to check the feasibility of running such a system in home routers or small business environments. We run experiments in our lab environment on an OpenWRT compliant Netgear router (model WNDR 3700v2) and an entity that hosts the UPS (MacBook Pro, Intel Core i5 and 8 GB of RAM). To check the system feasibility, the tests are performed by considering the total time spent (latency) by the osMUD Manager in:
\begin{itemize}
    \item retrieving the MUD file;
    \item verifying and storing the file locally at the router;
    \item processing the file and installing the rules.
\end{itemize}
These actions depict the setting time of the rules specified in a MUD file.

The tests consider router in its \textbf{booting stage} that represents the highest overloading phase, because of the aggregation of DHCP requests received from all connected devices (IoT and non-IoT). In case of MUD compliant IoT devices, the osMUD manager processes the MUD-URLs one by one to locate the MUD files. Hence, the file retrieving and processing is sequential. Additionally, to study the worst-case scenario, our tests examine the situation where all the IoT devices are turned on before the router boot process so that all the MUD files need to be retrieved at the same time. For testing purpose, the MUD file server provided by osMUD designers \footnote{https://mudfiles.nist.getyikes.com/} is used.

The osMUD manager, when a MUD compliant IoT device has administrator rules (UPS file), has to repeat the same operations listed above after installing the MUD file rules. It is worth noting that these UPS files are not defined for all the MUD compliant device, but only for devices for which is considered appropriate by an administrator (in our test it is simulated with a random number).

\begin{figure}%
	\centering
	\includegraphics[width=0.4\textwidth]{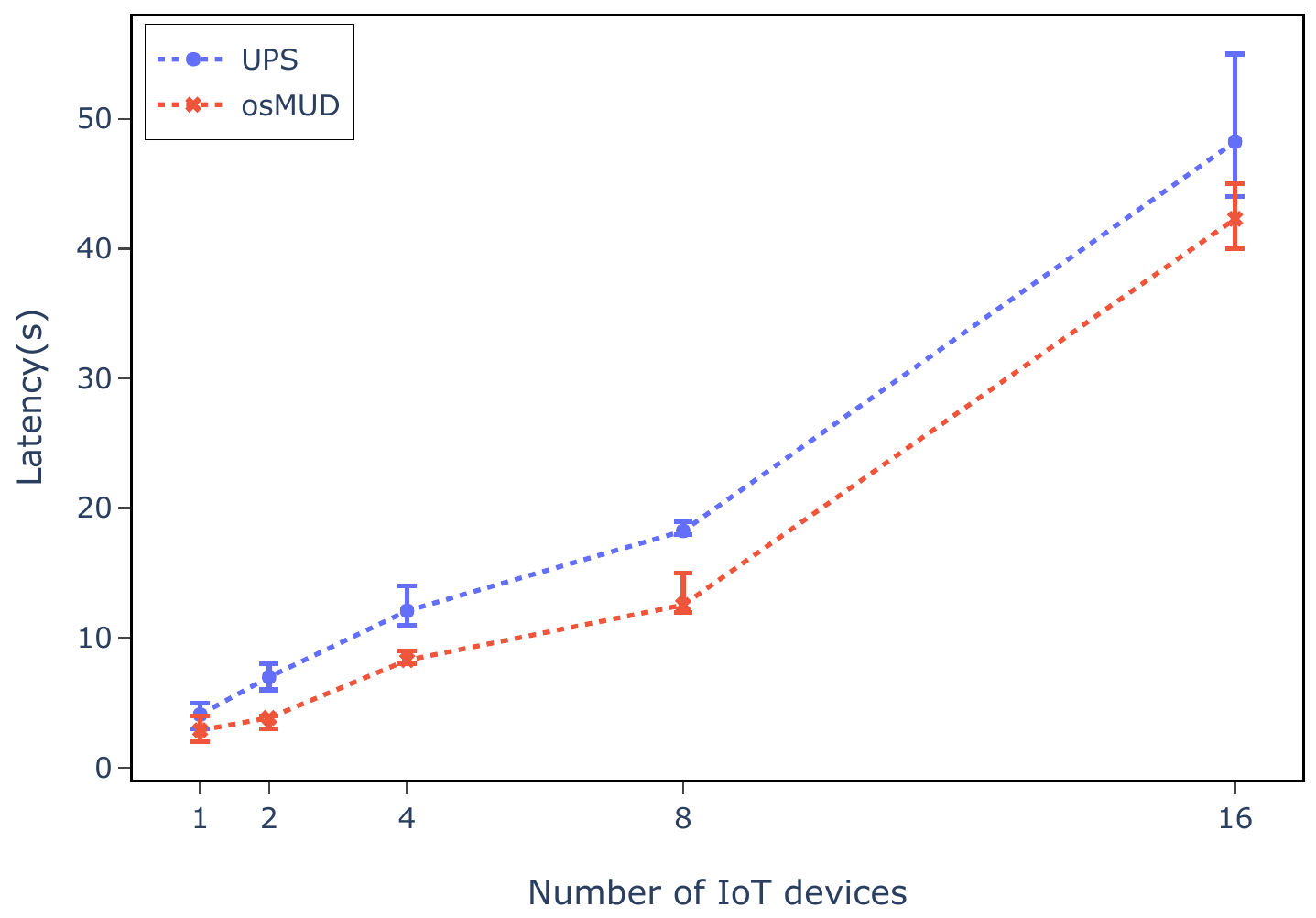}
	\caption{Access Control List (ACL) Rule setting time with osMUD and osMUD with UPS.}
	\label{fig:aclTime}
\end{figure}

The Fig. \ref{fig:aclTime} illustrates the performance of the osMUD manager in classical (without UPS files) and administrator (with UPS files) environment. Both the tests have been repeated 20 times with a file size range between 2-6K. In the test settings, the UPS file requests and MUD file requests are set as: (1,1), (2,2), (2,4), (3,8), (6,16). The numbers in the brackets (a,b) indicate:  a = number of MUD compliant devices (random), which request for UPS file; b = number of MUD compliant devices.

First of all, by considering only the one device scenario, the performance registered by the osMUD manager in the presence of UPS file is slightly worse than those achieved by the classical MUD environment, as a result of an additional request done to the UPS. Nevertheless, the results obtained are better than expected; the latency average in osMUD with the UPS file is 4.1 seconds compared with 2.9 seconds of osMUD in classical cases, which means in average only 1 second worse. One of the reasons that bring out this result comes from the evidence that requests do not leave the local network, which implies a reduction in the file download time. These results are promising and motivate the usage of a User Policy Server in a MUD deployment.

Secondly, in the ideal environment, the time spent on doing these operations is not linear, as a result of not stable connectivity. However, the time is not exponential, so the number of devices do not carry excessive weight on the performance. This is not true for the administrator environment, in which the time depends on the number of devices requires additional requests. For example, the couple (6,16) reached a spike of 55 seconds, compared with 45 seconds of the ideal environment.

In conclusion, in both cases with 16 devices, the time exceeds more than twice the time spent by 8 devices. The evaluation leads us to think that when the number of devices is over a certain threshold, the router starts to slow its performance, for example, in the case illustrated in Fig. \ref{fig:aclTime} the threshold value can be defined as 8 devices, as a result of the considerable growth reached after this number. The circumstance described represents the worst condition that could occur in case of router failure or unexpected reboot of it. In a steady condition in which at most 2 devices at a time are added, the performance registered are acceptable in both the environment even if both the devices requests an additional UPS file.

\section{Related Work}\label{related}
The use of MUD as an isolation-based defensive mechanism to restrict traffic generated from the \iot devices is still in its early phase. Therefore, only a few deployment scenarios and proof-of-concept (PoC) implementations currently exist.
Andalibi et al.~\cite{Andalibi2019} proposed extended MUD functionality to include traffic patterns - such as message size, peak-rate and maximum message transmissions within a time-period and proposed to implement those filtering at the fog network firewall. However, the paper does not provide any implementation. Also, it does not provide further details related to how these traffic control lists will be enforced at the fog networks and how this solution will be scalable in a large organization.   Yadav et al.~\cite{Yadav2019a} have proposed the scenario of combining user control policies with MUD to provide fine-grained control of IoT data. This work lacks to provide how to use user-based policies with MUD to control messages directly exchanged with external services.  Ranganathan et al.~\cite{ranganathan2019soft} presented a scalable implementation of the MUD standard on OpenFlow-enabled Software Defined Networking switches and  Hamza et al.~ \cite{Hamza2018a, Hamza2019 } developed machine learning methods to detect volumetric attacks such as DoS, reflective TCP/UDP/ICMP flooding, and ARP spoofing to IoT devices in MUD compliant traffic. Feraudo et al.~\cite{Feraudo2020} presented and evaluated Federated Learning in the MUD compliant Edge Network and showed a use-case of User Policy Server (UPS). Gracia et al.~\cite{Garcia2019} presented a proactive security approach managing secured and automated deployment of IoT devices by using an SDN-enabled security architecture in Industrial IoT settings.  Matheu et al.~\cite{Matheu2020} presented an extended  MUD model with a flexible policy language to express additional aspects, such as data privacy, channel protection, and resource authorization. Afek et al.~\cite{Afek2019}  presented a system built upon the MUD specification to provide IoT security through a VNF deployed within the ISP.  
\section{Conclusions and Future works}\label{conclusion}
The MUD specification presents some limits from security and architecture perspectives. There exist some behaviours which can be only defined by local policy, even when using components that are fully MUD compliant. If the default policy provided by the manufacturer is not sufficient or too restrictive for the network standard, user actions are necessary to configure the device according to a different and desired policy. Thus, our work explores the existing MUD implementations and tries to find the most convenient deployment for the general user. Furthermore, our work implements the UPS web server that provides a user-friendly interface needed to interact with the MUD components to modify their default settings when required. The behaviour defined allows building whitelist rules more suitable for the network in which MUD is deployed.

Our future directions will be focused on the extension of the User Policy Server functionalities and on solving implementation inconsistencies concerns, to guarantee a higher level of security. The work we presented in this paper provides useful insight, and we hope the research and development community get benefit from this. Furthermore, this work opens up more discussions before we start adopting these solutions in real deployment scenarios.

{
  \bibliographystyle{IEEEtran}
  \bibliography{iot} 
}

\end{document}